\documentclass{amsart}

\usepackage[pdftex]{graphicx}
\usepackage{amsmath,amssymb}
\DeclareMathAlphabet{\mathpzc}{OT1}{pzc}{m}{it}

\newcommand{\R}{\mathbb{R}}
\newcommand{\B}{\mathcal{B}}
\newcommand{\Lebesgue}{\mathcal{L}}
\newcommand{\lint}{\int\limits}
\newcommand{\ie}{\textit{i.e.}}

\newtheorem{theorem}{Theorem}

\graphicspath{{./}{figures/}}

\begin{document}

\title[Modeling self-organization in pedestrians and animal groups]{Modeling self-organization in pedestrians and animal groups from macroscopic and microscopic viewpoints}

\author{Emiliano Cristiani}
\address{CEMSAC - Universit\`a degli Studi di Salerno, Fisciano (SA), Italy and IAC-CNR, Rome, Italy}
\email{emiliano.cristiani@gmail.com}

\author{Benedetto Piccoli}
\address{IAC-CNR, Rome, Italy}
\email{b.piccoli@iac.cnr.it}

\author{Andrea Tosin}
\address{Department of Mathematics, Politecnico di Torino, Turin, Italy}
\email{andrea.tosin@polito.it}
\thanks{A. Tosin  acknowledges the support of a fellowship by the National Institute for Advanced Mathematics (INdAM) and the ``Compagnia di San Paolo'' foundation.}

\begin{abstract}
This paper is concerned with mathematical modeling of intelligent systems, such as human crowds and animal groups. In particular, the focus is on the emergence of different self-organized patterns from non-locality and anisotropy of the interactions among individuals. A mathematical technique by time-evolving measures is introduced to deal with both macroscopic and microscopic scales within a unified modeling framework. Then self-organization issues are investigated and numerically reproduced at the proper scale, according to the kind of agents under consideration.
\end{abstract}


\maketitle

\section{Self-organization in many-particle systems}
\label{sect:self-org}
One of the most outstanding expressions of intelligence in nonclassical physical systems, such as human crowds or animal groups, is their self-organization ability. Self-organization means that the individuals composing the system can give rise to complex patterns without using intercommunication as an essential mechanism.

\begin{figure}[t]
\begin{center}
\begin{minipage}[b]{0.4\textwidth}
\begin{center}
\includegraphics[width=\textwidth,height=6.85cm,clip]{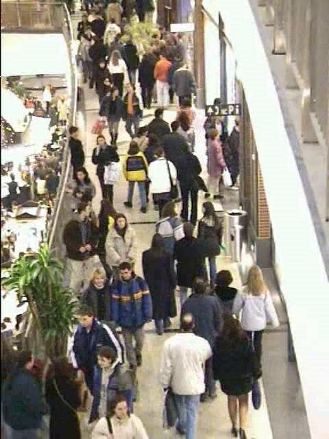} \\ (a) \\
\includegraphics[width=\textwidth,clip]{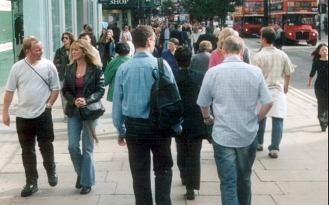} \\ (b)
\end{center}
\end{minipage}
\qquad
\begin{minipage}[b]{0.4\textwidth}
\begin{center}
\includegraphics[width=\textwidth,clip]{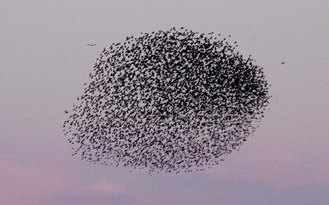} \\ (c) \\
\includegraphics[width=\textwidth,clip]{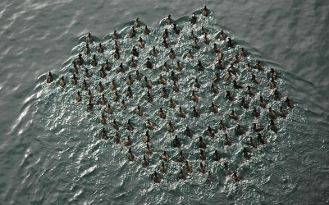} \\ (d) \\
\includegraphics[width=\textwidth,clip]{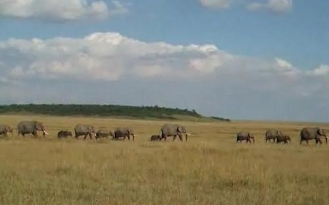} \\ (e)
\end{center}
\end{minipage}
\end{center}
\caption{Self-organizations. (a), (b): oppositely walking lanes of pedestrians. (c): three-dimensional globular cluster of starlings. (d): two-dimensional crystal-like globular cluster of surf scoters. (e): a line of migrating African elephants. Reproduction of these pictures with kind permission of the respective copyright holders. Credits are in the acknowledgments at the end of the paper.}
\label{fig:photos}
\end{figure}

For instance, in normal conditions pedestrians are known to arrange in specific patterns, chiefly lanes (cf. Fig. \ref{fig:photos}a-b), as demonstrated by many experimental investigations \cite{HeFaMoVi,HeJo,HeMoFaBo,HoDa1,HoDaBo}. Lane formation may be fostered by a suitable setup of the space, as reported in \cite{HeFaMoVi,HeMoFaBo}: a test performed in a tunnel connecting two subway stations in Budapest showed that a series of columns, placed in the middle of the walkway, induce pedestrians to organize in two oppositely walking lanes, preventing each of them to expand up to the full width of the corridor. More in general, lanes form also spontaneously, \ie, without the need for being triggered by environmental factors, provided the density of pedestrians is sufficiently large \cite{HeJo}. This is particularly evident if one considers the case of two groups of people, walking in opposite directions, which meet and cross (see also \cite{HoDa1}).

Grouping and self-organization are well known and largely observed also in animals, see for example \cite{KrRu}. These phenomena are in fact ubiquitous, ranging from bird flocks in the sky to migrating lobsters on the sea floor. Many papers on this subject (see, among others, \cite{chate2008mcm,couzin2005ela,couzin2002cma,cucker2007ebf,gregoire2003mst,gueron1996dhi,huth1992smf,kunz2003afs,
li2008mms,vicsek1989ntp,warburton1991tdm}) proved by means of numerical simulations that few simple rules adopted by each animal can give rise to a complex organization of the whole group. 
Patterns commonly seen in nature are (see \cite{heppner1974aff} for the nomenclature): globular clusters (e.g., starlings, Fig. \ref{fig:photos}c, surf scoters, Fig. \ref{fig:photos}d), extended and front clusters (e.g., wildebeests, antelopes, pigeons), lines (e.g., penguins, lobsters, elephants, Fig. \ref{fig:photos}e), Vees, Jays, and echelons (e.g., geese). Motivations and convenience of these patterns are still under investigations, especially for Vees.

It is commonly agreed that self-organization is the result of elementary actions that each subject performs to fulfill specific wills. Concerning pedestrians, the following basic guidelines can be identified:
\begin{itemize}
\item The will to reach specific targets, e.g., an exit or a meeting point, which drives pedestrians along preferential paths, determined mainly by the geometry and the spatial arrangement of the walking area. Unlike animals, pedestrians experience strong interactions with the environment, because they usually move in highly structured spaces scattered with all sorts of obstacles.

\item The will to not stay too close to one another, with a preference for uncrowded areas (\emph{repulsion} from other individuals). Pedestrians may agree to deviate from their preferred path, looking for free surrounding room.
\end{itemize}
In addition, it is reasonable to believe that occasionally also a mild form of cohesion occurs, which translates the tendency of pedestrians to not remain isolated. This happens, for instance, in those groups whose individuals share specific relationships, such as groups of tourists in guided tours.

For animals, the basic guidelines can instead be outlined as follows:
\begin{itemize}
\item The will to not stay too close to one another, in order to avoid collisions (\emph{repulsion} from other individuals).

\item The will to not remain isolated (group \emph{cohesion}). Grouping is in fact advantageous for many reasons, such as predator avoidance or food search, see \cite{KrRu}.
\end{itemize}
Like pedestrians, also animals may want to reach some specific destinations (e.g., when they migrate). However, the direction of motion toward targets is usually almost constant for quite a long time, because animals move in large and basically obstacle-free environments. Therefore, this aspect plays a minor role in the description of their self-organization, corresponding to a simple translation of the center of mass of the whole group.

Self-organization can be broken under particular circumstances entailing a dramatic change in the basic interaction rules discussed above. For pedestrians, an illuminating example is panic, when individuals tend to cram toward a common target (e.g., an escape) instead of seeking the less congested paths. Lane formation is then ruled out and less organized patterns emerge in the crowd, probably in consequence of a strong simplification of the interaction rules. Nevertheless, the arrangement of the environment may help restore, at least partially, the normal order of the flow. A typical example is that of an obstacle placed in front of an exit, which in some situations (presumably under panic conditions) may improve the flow of people through the exit itself, provided shape, dimensions, and position of the obstacle are accurately studied. This is a variant of the so-called Braess' paradox \cite{hughes2003fhc}, which states that a condition intuitively expected to lead to a worse situation may instead give rise to better outcomes. In animals, the changes in the behavior are even more evident. External conditions (e.g., presence of predators, weather), group tasks (feeding, exploring) or group speed can modify the interaction rules, leading to great modifications in the resulting patterns of the group. The environment usually does not interfere much, and pattern formation is mainly due to interactions among the individuals.

In this paper we are concerned with mathematical modeling of the above-discussed systems, with the specific aim of describing the spontaneous emergence of self-organization. In particular, we focus on two basic characteristics of the interactions among the individuals, namely non-locality in space and anisotropy. The inclusion of these factors makes our models able to explain the differences observed in self-organization and pattern formation of various groups of agents in terms of different visual fields and sensing zones.

At the same time, we introduce a modeling framework based on the measure theory, that allows for a unified formulation of macroscopic and microscopic models in any space dimension, and for a convenient numerics. This enables us to investigate both macroscopic self-organization, typical of large crowds of pedestrians, and microscopic self-organization, more specific of animal groups, using common modeling principles and tools.

In more detail, the paper is organized as follows. After this introduction, Sect. \ref{sect:class-vs-int} proposes a comparison between classical and intelligent particles, and highlights the main differences of the latter with respect to more standard systems dealt with by classical physics. In view of these considerations, Sect. \ref{sect:new_mech} develops some preliminary modeling strategies to enhance intelligence as a distinctive feature of the systems at hand. These serve as guidelines for Sect. \ref{sect:model}, where the modeling technique by time-evolving measures is introduced and specific macroscopic and microscopic models are detailed. Numerical tests on the ability of these models to reproduce spontaneous self-organizing patterns for both human crowds and animal groups are performed and commented in Sect. \ref{sect:numtest}. Finally, Sect. \ref{sect:conclusions} draws some conclusions and briefly sketches research perspectives.

\section{Classical vs. intelligent particles}
\label{sect:class-vs-int}
A plethora of physical systems can be basically described as systems of interacting particles, think for instance of fluids, gases, and similar matters. Actually, also human crowds and animal groups are susceptible to this rough characterization, although it should be clear from the previous discussion that the particle analogy is only formal. Indeed, intelligence is what really makes the difference in this context: it gives rise to a decision-based dynamics determined by individual behavioral rules, rather than to a classical dynamics passively ruled by inertia. For this reason, human beings and animals are more properly characterized as Intelligent Particles (denoted IPs in the sequel for brevity). IPs can act directly on the system, rather than being passively subjected to the evolution itself, whence their ability to self-organize and to generate complex patterns.

In the following, we summarize the main differences between classical and intelligent particles, in order to outline the main novelties posed by intelligent systems with respect to more standard frameworks.

\subsubsection*{Robustness vs. Fragility.}
Classical particles are \emph{robust}, in the sense that they interact almost exclusively through collisions. For instance, gas particles change direction of motion and velocity only when hit by other particles, or possibly when they collide with the walls of the container in which the gas is stored. Conversely, IPs are \emph{fragile}, they try as much as possible to avoid mutual collisions as well as to steer clear of walls and obstacles scattered along their path.

\subsubsection*{Blindness \& Inertia vs. Vision \& Decision.}
Classical particles are \emph{blind}, indeed they have no information on the environment and on the distribution of the surrounding particles. Therefore, their dynamics is essentially ruled by \emph{inertia}, \ie, a passive response to the mechanical cues coming from the exterior. On the contrary, human beings and animals feature specific \emph{visual fields}, hence they can obtain information on the surrounding environment (e.g., presence of obstacles, of walls) and on the current distribution of (a certain number of) other agents. This information is then used to \emph{make decisions} on the future individual evolution, generally by following some ordinary behavioral rules proper of the kind of agent under consideration.

\subsubsection*{Local vs. Nonlocal Interactions.}
Collisions among, say, gas particles require the latter to be sufficiently close to hit each other. Due to the typical size of gas molecules, by far much smaller than the environments where they flow, this has been classically understood as if the colliding particles were occupying the same spatial position at the moment of the impact (see e.g., \cite{MR1942465} and the references therein), hence assuming the conceptual approximation of \emph{local}, \ie, pointwise, interactions. Conversely, IPs do not interact mechanically, rather they are influenced by the presence of other individuals or objects a certain distance away, that they want either to approach or to avoid. The resulting interaction is thus \emph{nonlocal}, because the agents do not need to be in contact to interact. 

We notice that there exist basically two types of nonlocal interaction: a \emph{metric} one, such that each agent is influenced by \emph{all} other agents located at a distance less than a given threshold; and a \emph{topological} one, such that each agent is influenced by a \emph{given number} of other agents, no matter how far they are\footnote{It is worth noting that in the biological literature authors often regard both metric and topological interactions as \emph{local} models, in contrast to \emph{nonlocal} models in which every agent interacts with all other group mates.}. Topological interactions have been used in several models, e.g., \cite{huth1992smf,inada2002ofm,parrish2002sof,warburton1991tdm}, this idea being supported by experimental investigations like, for example, \cite{aoki1980asb} on fish schools. Other models \cite{couzin2005ela,couzin2002cma,gueron1996dhi,kunz2003afs} do not include this feature in favor of a purely metric approach. However, recent results \cite{ballerini2008ira} show that starlings interact topologically with other six/seven group mates. It is reasonable to believe that topological interactions are common in nature, even in pedestrians, since they are mainly due to a limited capacity in processing the information. On the other hand, a purely topological interaction in a wide domain is not realistic, for an individual may not be concerned with very far mates (e.g., because it does not see them at all).

\subsubsection*{Isotropy vs. Anisotropy.}
Another striking difference concerning the way in which classical and intelligent particles interact is that the first are \emph{isotropic}, meaning that they are equally affected by mechanical cues coming from all directions. IPs are instead \emph{anisotropic}, \ie, they are sensitive to stimuli coming from specific directions. This is partly related to the width of their visual field, which in pedestrians coincides with the half-space in front of them, whereas in animals covers often almost all the surrounding space. More precisely, the visual field must be intended as an upper bound for the regions in which cohesion and repulsion are active. Indeed, repulsion can be expected to be mainly felt against the IPs ahead rather than against those on the side or behind, especially in running people or fast-moving animals. Cohesion instead is active toward the group mates in front, if the goal is to follow the head of the group, or in every direction, if the goal is the unity of the group.

The ideas of limited visual field, expressed in terms of blind rear zone, and of anisotropic sensing are quite common in the biological literature, see for instance \cite{couzin2002cma,gueron1996dhi,hemelrijk2008sos,huth1992smf,inada2002ofm,kunz2003afs,lukeman2009cmm}. However, they mainly play the role of passive features of the agents, and are not regarded as active features able to influence the shape of the group.

\subsubsection*{Energy \& Entropy vs. Self-organization.}
Particle collisions allow to describe the mechanics of classical systems by invoking the balance of linear momentum, as well as energy and entropy principles. A straightforward extension of these ideas to intelligent systems is not possible, because the ability of IPs to make subjective decisions continuously puts and removes energy in and from the system, in hardly quantifiable amounts. Entropy criteria may in turn be questioned, because entropy is classically related to the equiprobability of the states of a system, whereas self-organization promotes specific configurations of the IPs. The idea that, in living matter, energy and entropy play a somehow atypical role was already suggested by Schr\"oedinger \cite{schroedinger1967wil} in the Sixties, who formalized it through the concept of \emph{negative entropy}.

\vskip0.3cm
It is useful, at this point, to consider a very borderline system with respect to the ``classical vs. intelligent'' dichotomy, namely vehicular traffic. Traffic is, in principle, an intelligent mechanical system, because the intrinsic mechanics of the vehicles is tempered by the presence of the drivers, who determine a non strictly mechanical behavior of the cars. However, many successful models, relying on the fluid dynamical analogy of the flow of vehicles along a road, have been proposed in the literature (see \cite{HoBo,piccoli2009vtr} for comprehensive reviews on microscopic, kinetic, and macroscopic models). This approach has been possible, despite the nonstandard nature of the system at hand, because traffic is essentially one-dimensional, so that vehicles regulate their speed only. Such a dimensional constraint reduces significantly the possibility of self-organization and pattern formation, therefore fluid dynamical modeling, entropy reasonings, and finally the development of models and theories based on nonlinear conservation laws are feasible.

\section{What mechanics for intelligent systems?}	\label{sect:new_mech}
The discussion of the previous sections has highlighted a crucial feature of intelligent systems: their dynamics is only partially determined by classical (pseudo-)mechanical cues, therefore, in principle, it is not fully describable within the classical Newtonian framework of point mechanics. Concerning this, we note that in the systems we are considering impulsive forces are often present. For example, an animal which starts moving reaches in a very short time its final velocity, which then remains constant for a while. Including in the models such impulsive forces through a Newtonian approach is quite difficult and, after all, needless.

The dynamics of intelligent systems calls for a paradigm which enhances intelligence as a distinctive feature, rather than recovering it as a by-product of other principles. In this respect, let us consider that, from the mechanical/dynamical point of view, intelligence might be regarded as the ability of the agents to control their velocity \emph{actively}, \ie, without the need for inertial accelerations caused by external actions. In order to take this aspect into account, we suggest to split the velocity $v$ of the agents in two basic contributions:
\begin{equation}
	v = w+\nu,
	\label{eq:velocity}
\end{equation}
to each of which there correspond specific modeling strategies. In more detail:
\begin{itemize}
\item $w$ is the \emph{external velocity}, \ie, the component of the total velocity depending essentially on actions exerted on the agents by the external environment. For example, in human crowds these may be the repulsion produced by obstacles and walls or the attraction produced by targets (e.g., doors, displays, other people). In animal groups this effect is instead greatly reduced, because animals move in much less structured and generally clear environments. For the external velocity, a Newtonian-like modeling is feasible, hence $w$ can be deduced from inertial reasonings up to a careful identification of the external actions. 

\item $\nu$ is the \emph{intelligent velocity}, \ie, the component of the total velocity that the individuals control actively. The intelligent velocity is determined by the behavioral rules that the agents comply to, and by the decisions that they consequently make, therefore it need not be Newtonian. For instance, pedestrians and animals set up the intelligent velocity according to the occupancy of the surrounding space, so as to steer clear of congested areas while possibly preserving the compactness of the group. 
\end{itemize}

A convenient manner to model $\nu$ is by an \emph{equation of state} of the form
\begin{equation*}
	\nu(t,\,x)=f[\mathcal{Q}](t,\,x),
	\label{eq:nu}
\end{equation*}
where $t$, $x$ are time and space variables, respectively, $\mathcal{Q}$ comprises all state variables which can contribute to the determination of $\nu$ (e.g., the distribution of the agents in a suitable neighborhood, the metric and topological structure of the sensing zone), and $f$ is a functional relationship.

Many microscopic models of animal groups available in literature actually fit into such framework, and the same is true for macroscopic models using hyperbolic partial differential equations, see e.g., \cite{edelsteinkeshet2001mms,MR2257718} and references therein. The forces do not appear explicitly, the velocity being expressed directly as a function of the density of the animals, usually in a nonlocal way. On the other hand, different models at both the microscopic and the kinetic scale rely instead, at least formally, on a more Newtonian-like framework, in the sense that they recover the dynamics of the system from generalized forces responsible for the acceleration of the agents \cite{carrillo2009afd,ha2009fpk,helbing1997awm}. Similar considerations apply also to models of human crowds, for an overview of which we refer to \cite{helbing2009pce}. In particular, macroscopic models derived by closing the mass conservation equation with suitable relations for the velocity follow the ideas outlined above, see e.g., \cite{MR2158218,colombo2009ens,MR2438214,helbing2006aac,hughes2002ctf,hughes2003fhc}. Other models resort instead more heavily to classical fluid dynamical analogies, indeed they supplement the mass conservation equation by a momentum balance equation invoking concepts like acceleration or generalized forces and pressures \cite{MR2438218}. In these models, it is in principle possible to introduce topological interactions among IPs, as well as anisotropic sensing zones. Unfortunately, models based on PDEs and nonlinear conservation laws suffer from an important drawback: their mathematical treatment and numerical implementation get rather complicated in dimensions greater than one. Indeed, in realistic cases of interest for applications one must deal with possibly punctured two-dimensional domains (the holes representing the obstacles), which requires to handle boundary conditions in a way that may not be immediately suited for hyperbolic equations. Therefore, we prefer to use an alternative modeling strategy, which, at the same time, enables us to develop a modeling framework capable to treat both the macroscopic and the microscopic scale within a unified theoretical formulation. Then we will show the ability of specific models, deduced from this framework, to describe the emergence of self-organizing behaviors at the proper scale, according to the kind of agents under consideration.

\section{Mathematical modeling by time-evolving measures}	\label{sect:model}
In this section we set up a modeling framework for the study of the behavior of IPs. In doing this, we resort to a discrete-time dynamics and describe the occupancy of the space by the agents at different times via a sequence of positive Radon measures. This approach is inspired by \cite{CaFaTi}, where the authors address \emph{rendez-vous} problems for multi-agent systems, and has already been investigated in recent works \cite{piccoli2009tem,piccoli2009pfb}. Here we show furthermore that also microscopic models for animal groups can be recast in the time-evolving measures framework. Therefore, we ultimately provide a unified modeling procedure to treat both macroscopic and microscopic intelligent systems.

Let us consider a domain $\Omega\subseteq\R^d$, which represents the area where IPs are located and move. In our discussion, $d$ is the dimension of the domain, from the physical point of view it may be $d=1,\,2,\,3$. A great advantage of our approach is that there are basically no differences in the theory for different dimensions. At every time instant $n\geq 0$, we define a Radon positive measure $\mu_n$ over the Borel $\sigma$-algebra $\B(\Omega)$, such that, for each measurable set $E\in\B(\Omega)$, the number $\mu_n(E)\geq 0$ measures the occupancy of the area $E$ by the IPs. In other words, the mapping $\mu_n$ provides the \emph{localization}, \ie, the distribution, of the agents in the domain $\Omega$ at time $n$. The evolution to the next time $n+1$ depends on the dynamics of the system, that we describe via a \emph{motion mapping} $\gamma_n:\Omega\to\Omega$:
\begin{equation}
	\gamma_n(x)=x+v_n(x)\Delta{t}, \qquad x\in\Omega.
	\label{eq:motion_mapping}
\end{equation}
In practice, in the time step $n\to n+1$ the point $x$ is advected to $\gamma_n(x)$ by the velocity field $v_n:\Omega\to\R^d$. The duration of the time step is $\Delta{t}>0$. Then, the measure $\mu_{n+1}$ is constructed by pushing $\mu_n$ forward with $\gamma_n$, that is $\mu_{n+1}=\gamma_n\#\mu_n$ or, more explicitly,
\begin{equation}
	\mu_{n+1}(E)=\mu_n(\gamma_n^{-1}(E)), \qquad \forall\,E\in\B(\Omega).
	\label{eq:push_forward}
\end{equation}
This corresponds to the simple idea that the number of IPs contained in $E$ at time $n+1$ coincides with their number at the previous time $n$ in the pre-image $\gamma_n^{-1}(E)$, therefore Eq. \eqref{eq:push_forward} expresses the conservation of the mass of IPs. By rewriting it in the equivalent form
\begin{equation*}
	\mu_{n+1}(E)-\mu_n(E)=-[\mu_n(\gamma_n^{-1}(E^c)\cap E)-\mu_n(E^c\cap\gamma_n^{-1}(E))],
	\label{eq:push_forward_2}
\end{equation*}
we recognize the formal statement of a conservation law: the left-hand side represents the variation of the measure of $E$ in a single time step, and the right-hand side the difference between the outgoing flux, \ie, the measure of the set of points belonging to $E$ and mapped outside $E$ by $\gamma_n$, and the incoming flux, \ie, the measure of the set of points not belonging to $E$ and mapped inside $E$ by $\gamma_n$.

Mathematical models are deduced from the structure \eqref{eq:motion_mapping}-\eqref{eq:push_forward} by prescribing the initial distribution of the IPs, that is the measure $\mu_0$, and by specifying the form of the velocity field $v_n$. According to the discussion of Sect. \ref{sect:new_mech}, we represent $v_n$ as the superposition of external and intelligent contributions:
\begin{equation*}
	v_n(x)=w(x)+\nu_n[\mu_n](x).
	\label{eq:vn}
\end{equation*}
The functional dependence of the intelligent velocity on the distribution of the agents, formally expressed by the square brackets in this formula, enables us to account for all of the factors having to do with the perception of the occupancy of the surrounding space from IPs (cf. Sect. \ref{sect:class-vs-int}). In more detail, we distinguish cohesive and repulsive effects in $\nu_n$, that we model as
\begin{align}
	\nu_n[\mu_n](x) &= \nu^c_n[\mu_n](x)+\nu^r_n[\mu_n](x) \notag \\
	&= \lint_{B_c(x)}F_c(y-x)\,d\mu_n(y)+\lint_{B_r(x)\setminus\{x\}}F_r\frac{y-x}{{\vert y-x\vert}^2}\,d\mu_n(y)
	\label{eq:nun-model}
\end{align}
for coefficients $F_c\geq 0$, $F_r\leq 0$. In this equation, $B_c(x),\,B_r(x)\subset\Omega$ are the \emph{zone of cohesion} and the \emph{zone of repulsion} of the agent $x$, respectively, which will be defined in the following.

The strength of the cohesive velocity $\nu^c_n$ is chosen to be growing linearly with the distance among IPs within the zone of cohesion $B_c(x)$. Cohesion  is mainly topological, \ie, it involves a predefined number of IPs. By consequence, for a fixed positive $p$, the zone of cohesion is chosen to satisfy $\mu_n(B_c(x))\leq p$, but its size may vary from point to point according to the distribution of the agents. However, a maximal size of $B_c(x)$ exists, say $s$, beyond which interactions among IPs are inhibited due to the distance (cf. Sect. \ref{sect:class-vs-int}), so that also $\Lebesgue^d(B_c(x))\leq s$ ($\Lebesgue^d$ being the Lebesgue measure on $\R^d$). In practice, $B_c(x)$ is adjusted dynamically under the constraints
\begin{equation}
	\mu_n(B_c(x))\leq p \quad \text{and} \quad \Lebesgue^d(B_c(x))\leq s.
	\label{eq:constraints-Bc}
\end{equation}

\begin{figure}[t]
\begin{center}
\includegraphics[width=0.4\textwidth,clip]{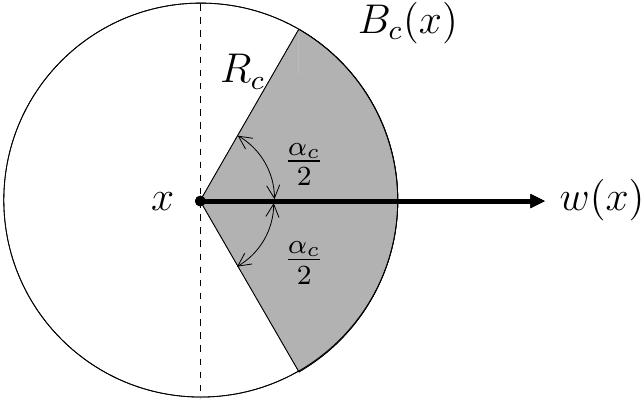}
\caption{Definition of the zone of cohesion by means of the velocity $w$, the topological radius $R_c$, and the span angle $\alpha_c\in(0,\,2\pi]$. Analogous considerations hold also for the zone of repulsion, up to considering a metric radius $R_r$ and an angle $\alpha_r$.}
\label{fig:Bc}
\end{center}
\end{figure}

Let us choose, for convenience, $B_c(x)$ to be a circular sector of the ball of radius $R_c>0$ (to be detemined), centered in $x$ and with central angle $\alpha_c\in(0,\,2\pi]$ (Fig. \ref{fig:Bc}). Other choices are possible, for instance in \cite{kunz2003afs} the authors use elliptical regions to take the body form of the agents into account.
Due to the symmetry of the body, we assume that the sensing domain of each IP is symmetric with respect to the main direction of motion fixed by the vector $w(x)$. Introducing the family of circular sectors \begin{equation*}
	S(x,R,\alpha)=\left\{y\in\Omega\,:\,\vert y-x\vert\leq R,\ \hat{r}(x,\,y)\cdot\hat{w}(x)\geq
		\cos{\frac{\alpha}{2}}\right\},
\end{equation*}
where $\hat{r}(x,\,y)$ and $\hat{w}(x)$ are the unit vectors in the directions of $y-x$ and $w(x)$, respectively, and setting
\begin{equation*}
	R_c=\max\{R\geq 0\,:\,\mu_n(S(x,R,\alpha_c))\leq p,\,\Lebesgue^d(S(x,R,\alpha_c))\leq s\},
\end{equation*}
we can easily define $B_c(x):=S(x,R_c,\alpha_c)$. Notice that the constraint on the Lebesgue measure of each $S$ basically amounts to fixing a metric upper bound $R_c^{\text{max}}$ to the maximum radius allowed for the zone of cohesion $B_c(x)$. This formalizes the topological cohesion complemented with a metric cut-off observed in \cite{ballerini2008ira} (see also \cite{gregoire2003mst}).

Conversely, we take the strength of the repulsive velocity $\nu^r_n$ proportional to the inverse of the distance among IPs within the zone of repulsion $B_r(x)$. Repulsion is mainly metric, as each IP simply tries to maintain a minimum distance between itself and other IPs. Hence $B_r(x)$ has a fixed size $\Lebesgue^d(B_r(x))$, while its measure $\mu_n(B_r(x))$ may vary from point to point according to the crowding of the space. Assuming again for convenience that $B_r(x)$ is a circular sector of the ball with center in $x$ and (fixed) radius $R_r>0$, we simply have $B_r(x):=S(x,R_r,\alpha_r)$, $\alpha_r\in(0,\,2\pi]$ being the angular span of the zone of repulsion.

Finally, the external velocity $w$ depends essentially on the interactions of the IPs with the environment. Instead, it is independent of the distribution of the agents in the domain, because it represents the velocity that an isolated agent would set to reach its targets. Due to the strong differences in the structure of the typical environments where pedestrians and animals move, we refrain from giving here a general structure of $w$. We will detail it in the next subsections, with reference to specific applications.

\subsection{Macroscopic models}	\label{sect:model-macro}
Macroscopic models are useful to study emergent self-organizing behaviors on large scales. This is the case of human crowds, in which self-or\-ga\-ni\-za\-tion entails the formation of patterns that are clearly visible only when the density of people is sufficiently high. In addition, these models are particularly handy to face the technical complexity of pedestrian flows in structured environments scattered with obstacles, possibly also in connection with safety and optimization issues.

Macroscopic models are obtained from the modeling framework outlined above via the continuum hypothesis, which amounts to saying that the measure $\mu_n$ is absolutely continuous with respect to the Lebesgue measure $\Lebesgue^d$ ($\mu_n\ll\Lebesgue^d$). Therefore, one can introduce a non-negative function $\rho_n\in L^1(\Omega)$ such that $d\mu_n=\rho_n\,dx$, and speak of density of IPs. The link with the measure $\mu_n$ is explicitly stated as
$$ \mu_n(E)=\lint_E\rho_n(x)\,dx, \qquad \forall\,E\in\B(\Omega). $$
Starting from an initial measure $\mu_0\ll\Lebesgue^d$, existence of the density for all times $n$ is provided by the following result proved in \cite{piccoli2009tem}:
\begin{theorem}
For all $n>0$, let a constant $C_n>0$ exist such that
$$ \Lebesgue^d(\gamma_n^{-1}(E))\leq C_n\Lebesgue^d(E), \qquad \forall\, E\in\B(\Omega). $$
If $\mu_0\ll\Lebesgue^d$ is nonnegative, then there is a unique sequence $\{\rho_n\}_{n\geq 1}\subset L^1(\Omega)$, $\rho_n\geq 0$ a.e. in $\Omega$, such that $\mu_n\ll\Lebesgue^d$ with $d\mu_n=\rho_n\,d\Lebesgue^d$ and $\|\rho_n\|_1=\|\rho_0\|_1$ for all $n>0$.

If in addition $\rho_0\in L^1(\Omega)\cap L^\infty(\Omega)$ then $\rho_n\in L^1(\Omega)\cap L^\infty(\Omega)$ as well, with $\|\rho_n\|_\infty\leq\left(\prod_{k=1}^{n}C_k\right)\|\rho_0\|_\infty$ for all $n>0$.
\label{theo:existence}
\end{theorem}

Under the continuum hypothesis, the intelligent velocity \eqref{eq:nun-model} specializes as
\begin{equation}
	\nu_n[\rho_n](x)=\lint_{B_c(x)}F_c(y-x)\rho_n(y)\,dy+
		\lint_{B_r(x)}F_r\frac{y-x}{{\vert y-x\vert}^2}\rho_n(y)\,dy.
	\label{eq:nun-continuum}
\end{equation}

Concerning the external velocity, to be definite we consider the application to pedestrians. In this case, it is convenient to model $w$ as a normalized potential flow $w=\nabla{u}/\vert\nabla{u}\vert$, which does not depend on the distribution $\rho_n$ of the people but only on the geometry of the domain (cf. Sect. \ref{sect:new_mech}), including the possible presence of obstacles. The function $u:\Omega\to\R$ is a scalar potential satisfying Laplace's equation $\Delta{u}=0$. Boundary conditions for this equation may be set to identify attractive (resp., repulsive) areas, such as doors or displays (resp., obstacle edges or perimeter walls). For instance, targets along the boundary $\partial\Omega$ of the domain may be characterized by the maximum potential, say $u=1$, whereas the remaining portion of the perimeter walls by the minimum potential, say $u=0$. On the internal boundaries, namely obstacle walls, the Neumann condition $\frac{\partial u}{\partial n}=0$ may be prescribed instead, which corresponds to zero normal component of the velocity $w$. Finally, the external velocity is generated by solving for $u$ the resulting stand-alone elliptic problem

\begin{figure}[t]
\begin{center}
\begin{minipage}[c]{0.4\textwidth}
\begin{center}
\includegraphics[width=\textwidth,clip]{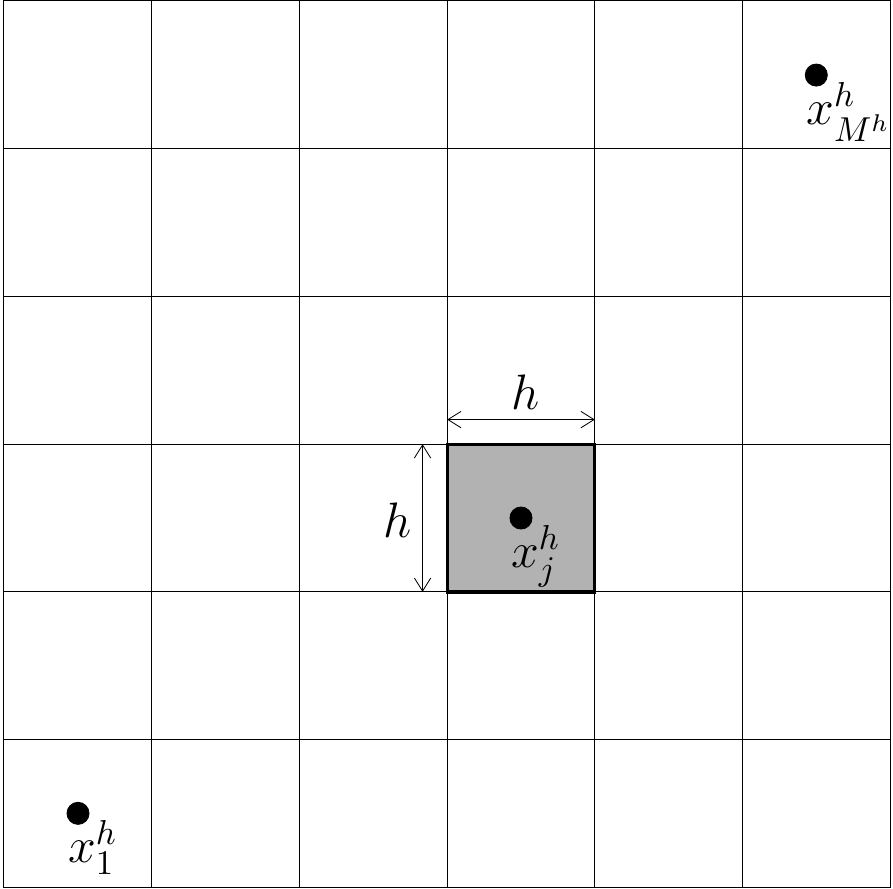}
\end{center}
\end{minipage}
\qquad\qquad
\begin{minipage}[c]{0.4\textwidth}
\begin{center}
\includegraphics[width=\textwidth,clip]{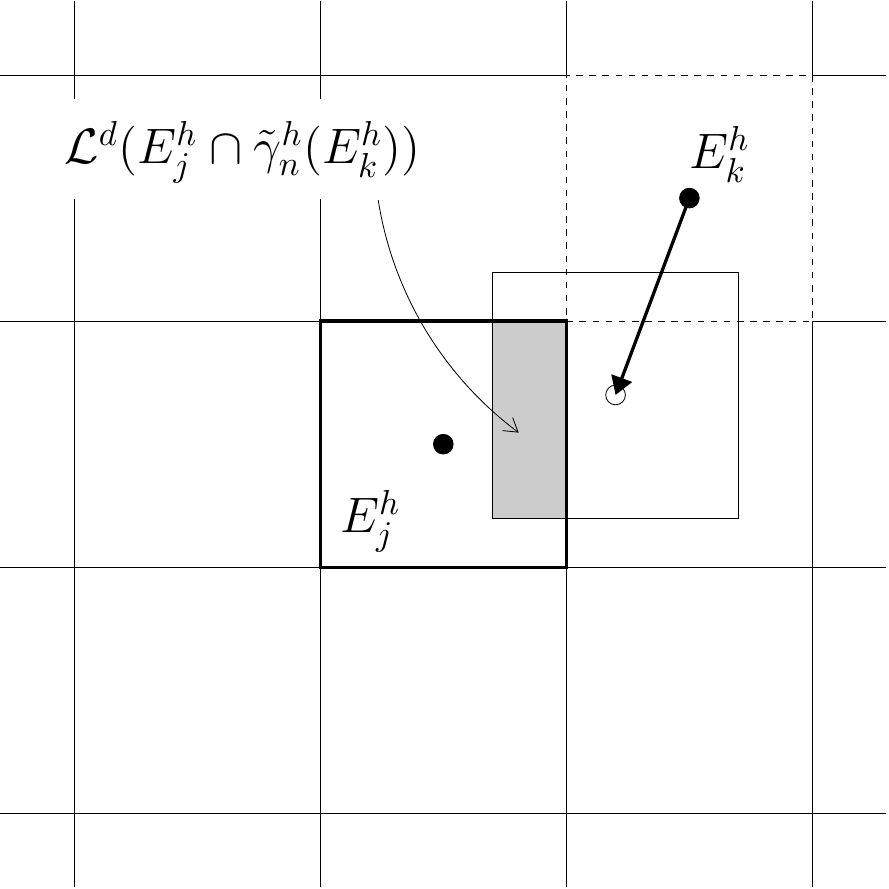}
\end{center}
\end{minipage}
\end{center}
\caption{Left: the grid $\{E_j^h\}_{j=1}^{M^h}$ in the domain $\Omega$, with the generic cell $E_j^h$ highlighted. Right: the coefficients of the numerical scheme \eqref{eq:numscheme} are determined as the (hyper)volumes of the overlapping parts of adjacent grid cells under the push-forward operated by the approximate motion mapping $\tilde{\gamma}^h_n$.}
\label{fig:grid}
\end{figure}

Equations are numerically solved on a finite volume-like partition of the domain $\Omega$, using an \emph{ad hoc} computational scheme deduced from the one suggested in \cite{piccoli2009pfb} with the inclusion of the new modeling features introduced here. Basically, a grid $\{E_j^h\}_{j=1}^{M^h}$ consisting of $M^h$ elements of characteristic size $h>0$ is introduced in $\Omega$ (see Fig. \ref{fig:grid}, left), and the density $\rho_n$ and the velocity $v_n$ are discretized as piecewise constant functions:
\begin{equation*}
	\left.
	\begin{array}{l}
		\rho_n(x)\approx\tilde{\rho}^h_n(x)\equiv \rho^h_{n,j}, \\[0.2cm]
		v_n[\rho_n](x)\approx\tilde{v}^h_n[\tilde{\rho}^h_n](x)\equiv v_n[\tilde{\rho}^h_n](x_j^h)
	\end{array}
	\right\} \quad
	\forall\,x\in E_j^h,
\end{equation*}
$x_j^h$ being a point of the cell $E_j^h$, for instance its center. This induces naturally a discretization of the motion mappings $\gamma_n$:
$$ \gamma_n(x)\approx\tilde{\gamma}^h_n(x)=x+\tilde{v}^h_n[\tilde{\rho}^h_n](x)\Delta{t}, $$
where it should be noticed that the $\tilde{\gamma}^h_n$'s act as piecewise translations on the grid $\{E_j^h\}_{j=1}^{M^h}$. After defining the new measures $d{\tilde\mu}^h_n:=\tilde{\rho}^h_n\,d\Lebesgue^d$, the scheme is obtained by imposing the push-forward $\tilde{\mu}^h_{n+1}=\tilde{\gamma}^h_n\#\tilde{\mu}^h_n$ on the grid cells $E_j^h$ only, \ie, $\tilde{\mu}^h_{n+1}(E_j^h)=\tilde{\mu}^h_n((\tilde{\gamma}^h_n)^{-1}(E_j^h))$ for all $j=1,\,\dots,\,M^h$, which gives (cf. Fig. \ref{fig:grid}, right):
\begin{equation}
	\rho^h_{n+1,j}=\frac{1}{\Lebesgue^d(E_j^h)}\sum_{k=1}^{M^h}\rho^h_{n,k}\Lebesgue^d(E_j^h\cap
		\tilde{\gamma}^h_n(E_k^h)).
	\label{eq:numscheme}
\end{equation}
If the $\gamma_n$'s are sufficiently smooth, and if the spatial resolution $h$ and the time step $\Delta{t}$ are linked by a CFL-like condition, this scheme turns out to be nicely behaved in terms of stability and localization error:
\begin{theorem}
Let $\gamma_n$ be a diffeomorphism and let $h,\,\Delta{t}>0$ satisfy
$$ \Delta{t}\max_{j=1,\,\dots,\,M_h}\vert v_n[\tilde{\rho}^h_n](x_j^h)\vert_2\leq h, $$
where $\vert\cdot\vert_2$ is the Euclidean norm in $\R^d$. Then:
\begin{enumerate}
\item [(i)] There exists a constant $C>0$ independent of $h$ such that
$$ \sum_{j=1}^{M^h}\vert\mu_{n+1}(E_j^h)-\tilde{\mu}^h_{n+1}(E_j^h)\vert\leq C
	\left(\|\rho_n-\tilde{\rho}^h_n\|_1+h\right). $$
\item [(ii)] For each $n>0$, there exists a constant $C_n>0$ independent of $h$ such that
$$ \max_{j=1,\,\dots,\,M^h}\vert\mu_{n+1}(E_j^h)-\tilde{\mu}^h_{n+1}(E_j^h)\vert\leq
	\max_{j=1,\,\dots,\,M^h}\vert\mu_0(E_j^h)-\tilde{\mu}^h_0(E_j^h)\vert+C_nh^d. $$
\end{enumerate}
\label{theo:errorest}
\end{theorem}
The proof of this result can be recovered again in \cite{piccoli2009tem}.

\medskip

The estimates of Theorem \ref{theo:errorest} rely essentially on the $L^1$ metric, because they assume the existence of the densities $\{\rho_n\}_{n\geq 1}$ for the measures $\{\mu_n\}_{n\geq 1}$. However, it can be questioned that such a metric is not the optimal one to evaluate the distance between measures, even when densities are available. This is particularly true in the application to pedestrian flows, where self-organization phenomena may give rise to measures concentrated in thin areas of the domain (though not singular in view of Theorem \ref{theo:existence}). Small errors in the localization of these measures would be roughly estimated by the $L^1$ distance of the corresponding densities, even when the corresponding distributions of the crowd are intuitively close.

\begin{figure}[t]
\begin{center}
\includegraphics[width=0.6\textwidth,clip]{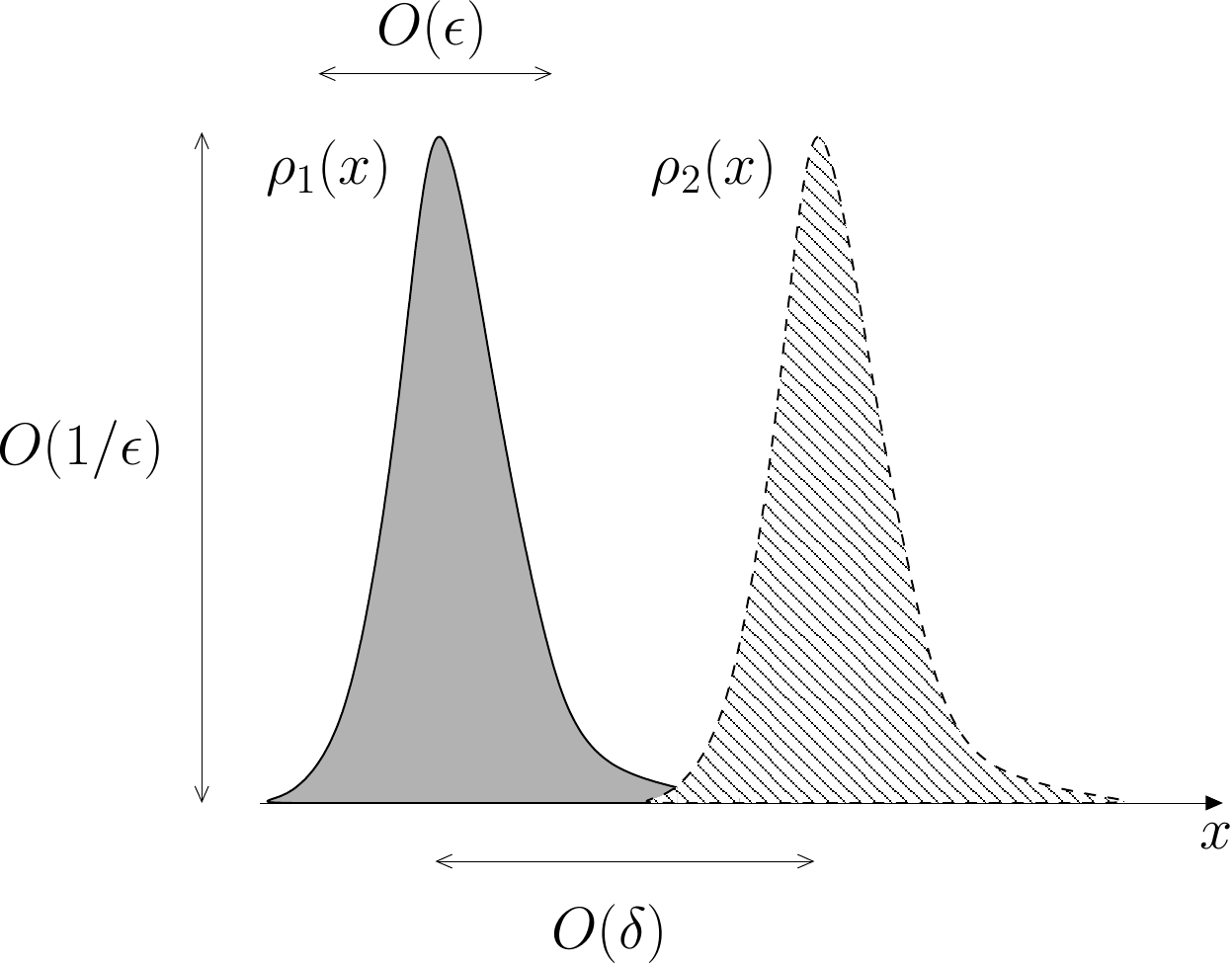}
\end{center}
\caption{Two measures $d\mu_i=\rho_i\,d\Lebesgue^d$, $i=1,\,2$, whose $L^1$ distance is $O(1)$ whereas Wasserstein distance is $O(\delta)$. The latter gives then a better estimate of the error produced by $\mu_2$ in the localization of $\mu_1$.}
\label{fig:Wass}
\end{figure}

To be definite, let us consider two measures $\mu_1,\,\mu_2$ with densities $\rho_1,\,\rho_2$, respectively, as illustrated in Fig. \ref{fig:Wass}. This kind of distributions may be seen, for instance, along the vertical cross section of crossing pedestrian flows (cf. Sect. \ref{ssubsect:lanes}), when oppositely walking lanes have fully emerged. Assume that $\|\rho_i\|_\infty=O(1/\epsilon)$, $\Lebesgue^d(\operatorname{supp}{\rho_i})=O(\epsilon)$, $i=1,\,2$, for some arbitrarily small $\epsilon>0$, whereas the average pointwise distance between $\rho_1$ and $\rho_2$ is $O(\delta)$ for $\delta>0$. Thus it is reasonable to expect a localization error of $\mu_2$ over $\mu_1$ of the order of $\delta$, but if $\delta$ is larger than $\epsilon$ and we use the $L^1$ norm we inevitably get $\|\rho_2-\rho_1\|_1=O(1)$, because most mass is concentrated in the non-overlapping parts of the supports of $\rho_1,\,\rho_2$.

A more correct way to measure the distance between $\mu_1,\,\mu_2$, which also better matches our intuition on what this distance should be, is offered by the \emph{Wasserstein metric}, that we can briefly introduce as follows. Let $(\Omega,\,\mathpzc{d})$ be a metric space for which every probability measure on $\Omega$ is a Radon measure. Then the Wasserstein distance between two probability measures $\mu_1,\,\mu_2$ is defined as
$$ W(\mu_1,\,\mu_2):=\inf_{T}\lint_\Omega\mathpzc{d}(x,\,T(x))\,d\mu_1(x), $$
the inf being taken among all transport maps from $\mu_1$ to $\mu_2$:
$$ \mu_2=T\#\mu_1, \qquad\text{\ie,}\qquad \mu_2(E)=\mu_1(T^{-1}(E)), \quad\forall\,E\in\B(E). $$
$W(\mu_1,\,\mu_2)$ is the best (\ie, the lowest) transportation cost to move the measure $\mu_1$ onto $\mu_2$ (and vice-versa). In our previous example, we would actually get $W(\mu_1,\,\mu_2)=O(\delta)$.

The discrete-time model \eqref{eq:push_forward} can be interpreted as an explicit Euler discretization in time of a gradient flow on the Wasserstein space, at least for sufficiently smooth motion mappings, cf. \cite{MR2401600,maury2009mcm-preprint,MR2459454}. Therefore the Wasserstein metric may be profitably used to further improve the theory with a more accurate error analysis. In particular, in future work we will try to obtain error estimates in space with respect to the Wasserstein distance, estimates in time being instead deducible from \cite{MR2401600}.

\subsection{Microscopic models}	\label{sect:model-micro}
Microscopic models are useful to study self-organization phenomena at small scale. They are common in biological literature, where a wealth of models were studied in order to understand grouping behavior in fish schools, bird flocks, mammals herds and bacteria aggregations. Microscopic models can be obtained from our time-evolving measures framework, removing the continuum hypothesis. Denoting by $x_j^n\in\Omega$ the position of the $j$-th IP at time $n$, the measure $\mu_n$ is now chosen to be the counting measure, \ie,
\begin{equation}
	\mu_n=\sum_{j=1}^{N}\delta_{x_j^n}, \qquad \mu_n(E)=\operatorname{card}{\{x_j^n\in E\}}
	\label{eq:dirac}
\end{equation}
where $N$ is the total number of agents and $\delta_x$ is the Dirac measure centered in $x$. Existence issues are now by far much easier than in the macroscopic case, because the measure $\mu_{n+1}$ can be constructed explicitly. Indeed, after a push-forward by $\gamma_n$ it is easily computed that
$$ \mu_{n+1}=\sum_{j=1}^{N}\delta_{\gamma_n(x_j^n)}, $$
hence at the next time step the new positions of the IPs are $x_j^{n+1}=\gamma_n(x_j^n)$. Finally, the resulting model turns out to be a classical agent-based model of the form
\begin{equation*}
	x_j^{n+1}=x_j^n+v_n(x_j^n)\Delta{t},
	\label{eq:agent-based}
\end{equation*}
which corresponds to the discrete-time version of the dynamical system $\dot{x}_j(t)=v(t,\,x_j(t))$ obtained by an explicit Euler scheme.

Inserting the measure \eqref{eq:dirac} in Eq. \eqref{eq:nun-model} yields the following expression of the intelligent velocity:
\begin{equation}
	\nu_n(x_j^n)=\sum_{x_k^n\in B_c(x_j^n)}F_c(x_k^n-x_j^n)+
		\sum_{x_k^n\in B_r(x_j^n)\setminus\{x_j^n\}}F_r\frac{x_k^n-x_j^n}{{\vert x_k^n-x_j^n\vert}^2}.
	\label{eq:nun-discrete}
\end{equation}

\begin{figure}[t]
\begin{center}
\includegraphics[width=0.4\textwidth]{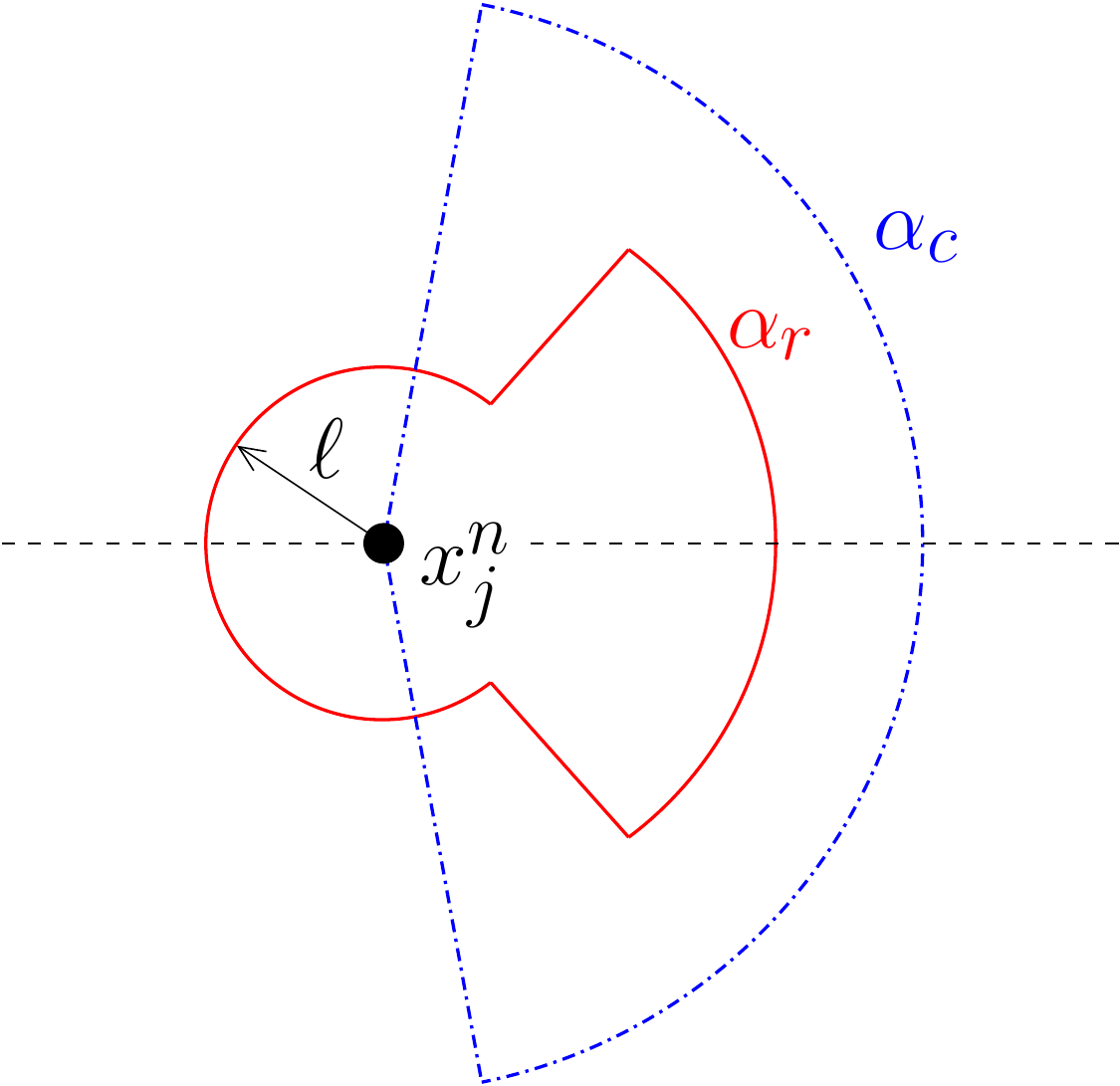}
\caption{At the microscopic scale, the body size $\ell$ of the IPs matters. Therefore, repulsion from the $j$-th agent is always active in a complete neighborhood of radius $\ell$, regardless of the radius $R_r$ and the angle $\alpha_r$. This, however, does not affect the zone of cohesion.}
\label{fig:zones-micro}
\end{center}
\end{figure}

Besides the formal derivation of Eq. \eqref{eq:nun-discrete} from Eq. \eqref{eq:nun-model}, we notice that at the microscopic scale the body size of the IPs matters. Indeed, the latter are not point agents, thus when two of them are closer than a characteristic distance $\ell>0$ they touch each other (cf. also \cite{maury2007hcc,maury2007mmf,maury2008mfc} about microscopic pedestrians). Therefore, we admit that within a small neighborhood of radius $\ell$ repulsion from a certain agent $x_j^n$ is active against all other agents, regardless of the angular span $\alpha_r$, and that the distance between two IPs is never smaller than $\ell$, cf. Fig. \ref{fig:zones-micro}. This also helps avoid singularities, \ie, non-integrability with respect to the Dirac measure, in the second sum of Eq. \eqref{eq:nun-discrete}.

Finally, we assume that the external velocity of the agents is constant (for instance, the vector $w=(1,\,0)$ for a rightward motion), as the environment is clear and we are mainly interested in the dynamics of the interactions among the IPs. This way, we can formally drop $w$ from Eq. \eqref{eq:vn} by a simple change of frame of reference.

\section{Numerical results}	\label{sect:numtest}
In this section we address some relevant case studies, which highlight the ability of our model to reproduce self-organizing patterns at both the macroscopic and the microscopic scale. In particular, we focus on pedestrians for macroscopic self-organization, studying the emergence of oppositely walking lanes in crossing flows, the spontaneous arrangement of people in a group in motion, and finally the dynamics in a crowded environment scattered with obstacles. 
Conversely, we study microscopic self-organization with specific reference to animals, for which a closer look is necessary in order to catch crystal-like structures and line formations.

\subsection{Macroscopic self-organization in pedestrians}	\label{sect:num-macro}
As recalled in Sect. \ref{sect:self-org}, pedestrians experience, in normal situations, a very mild cohesion among each other compared to repulsion. Hence, in most numerical tests below we set $F_c=0$ and consider the following overall velocity:
\begin{equation}
	v_n[\rho_n](x)=\frac{\nabla{u}(x)}{\vert\nabla{u}(x)\vert}+
		\lint_{B_r(x)}F_r\frac{y-x}{{\vert y-x\vert}^2}\rho_n(y)\,dy,
	\label{eq:vn-pedestrians}
\end{equation}
where $F_r<0$ is constant. Due to the ahead-behind asymmetry of pedestrians, their visual field covers a frontal area only. Therefore, we define the zone of repulsion $B_r(x)$ to be the half-ball of radius $R_r$ in the direction of the external velocity.

Unless otherwise stated, the values of the relevant parameters are, for all tests, $\alpha_r=\pi$, $R_r=0.1$, $F_r=-1$.

\begin{figure}[t]
\begin{center}
\includegraphics[width=0.3\textwidth,clip]{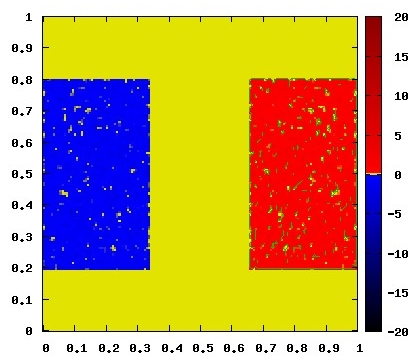}
\includegraphics[width=0.3\textwidth,clip]{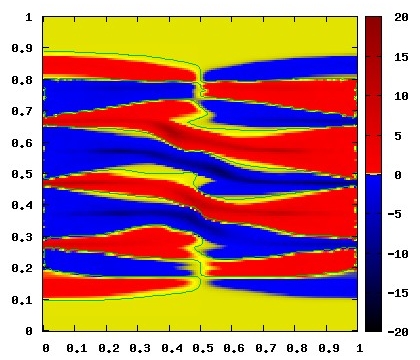}
\includegraphics[width=0.3\textwidth,clip]{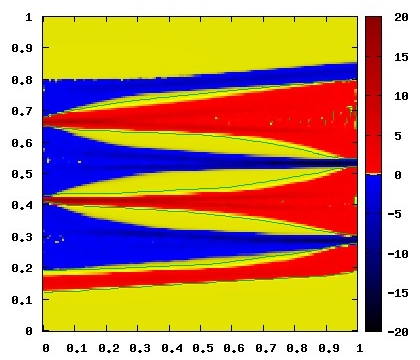} \\
\includegraphics[width=0.3\textwidth,clip]{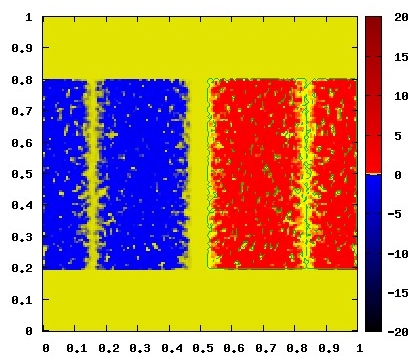}
\includegraphics[width=0.3\textwidth,clip]{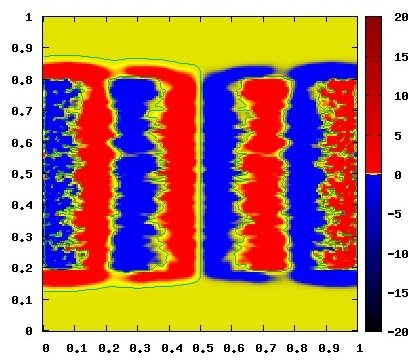}
\includegraphics[width=0.3\textwidth,clip]{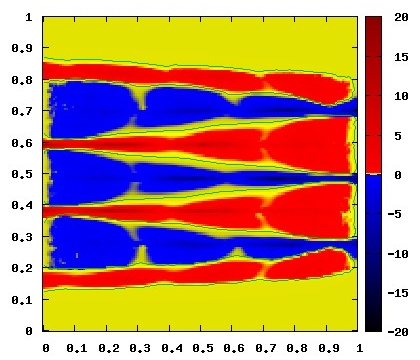}
\caption{Lane formation in crossing flows, with inhomogeneous uninterrupted (first row) and time-periodic (second row) injection of people from the boundaries of the domain. Pedestrians walking rightward are in blue, those walking leftward in red. Negative values of the density of the first ones are for graphical purposes only.}
\label{fig:crossflow}
\end{center}
\end{figure}

\subsubsection{Lane formation in crossing flows}
\label{ssubsect:lanes}
We consider two groups of pedestrians, of density $\rho_n^{(k)}\in L^1(\Omega)$, $k=1,\,2$, respectively, walking in opposite directions. This amounts to defining two measures $d\mu_n^{(k)}=\rho_n^{(k)}\,d\Lebesgue^2$, each of which evolves in time according to the push-forward \eqref{eq:push_forward}. However, the two evolutions are coupled, since we assume that, within each group, repulsion is oriented against the individuals of the opposite group. Specifically, this means that the intelligent velocities are such that:
\begin{equation*}
	\nu^{(1)}_n[\rho^{(2)}_n](x)=\lint_{B_r(x)}F_r\frac{y-x}{{\vert y-x\vert}^2}\rho^{(2)}_n(y)\,dy, 
	\label{eq:vin-crossflows_a}
\end{equation*}
\begin{equation*}
		\nu^{(2)}_n[\rho^{(1)}_n](x)=\lint_{B_r(x)}F_r\frac{y-x}{{\vert y-x\vert}^2}\rho^{(1)}_n(y)\,dy,
	\label{eq:vin-crossflows_b}
\end{equation*}
so that pedestrians try to steer clear of oppositely walking people and to gain instead room in their direction. The external velocities are the vectors $w^{(1)}=(1,\,0)$ and $w^{(2)}=(-1,\,0)$, thus groups walk rightward and leftward, respectively.

Figure \ref{fig:crossflow} shows that the model is able to account for lane formation when the two groups meet at the center of the domain and start to interact. In particular, alternate lanes fully emerge for both uninterrupted and time-periodic flow of people from the boundaries of the domain, and turn out to be a quite stable equilibrium configuration of the system. Such a configuration is however reached in different times, being in particular more delayed in the second case, as a consequence of different intermediate dynamics undertaken by the system. Notice also the spontaneous breaking of symmetry occurring between the two groups, which are instead specular at the beginning.

\begin{figure}[t]
\begin{center}
\begin{minipage}[c]{0.3\textwidth}
\centering
\includegraphics[width=\textwidth,clip]{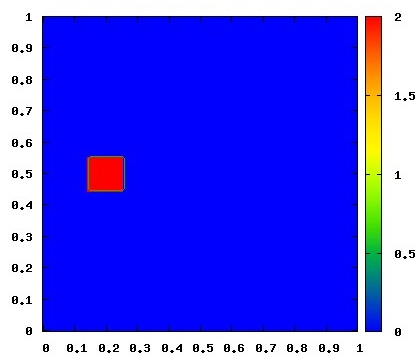} \\
(a)
\end{minipage}
\begin{minipage}[c]{0.3\textwidth}
\centering
\includegraphics[width=\textwidth,clip]{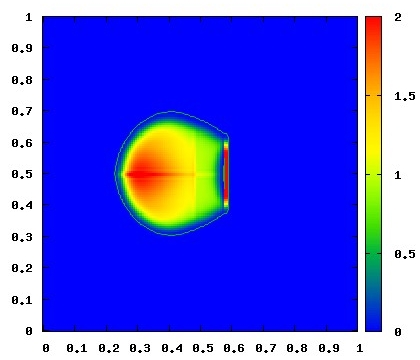} \\
(b)
\end{minipage}
\begin{minipage}[c]{0.3\textwidth}
\centering
\includegraphics[width=\textwidth,clip]{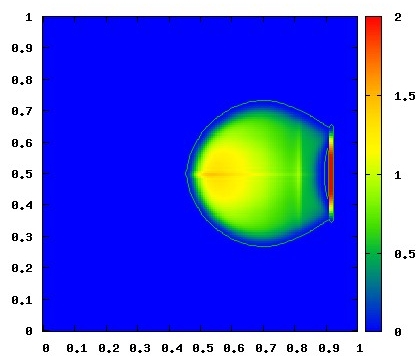} \\
(c)
\end{minipage}
\caption{Spontaneous arrangement of a crowd in motion. Starting from a compact cluster, the group expands and the density decreases. Leaders tend however to maintain the initial configuration.}
\label{fig:pedgroup}
\end{center}
\end{figure}

\subsubsection{Spontaneous arrangement of a crowd in motion}	\label{ssubsect:spontarrang}
Next we investigate the self-or\-ga\-ni\-za\-tion spontaneously emerging within a group of pedestrians in motion in a clear environment. The total velocity $v_n$ is now as in Eq. \eqref{eq:vn-pedestrians}, with an external velocity simply given by $w=(1,\,0)$ (the group is walking rightward). The group is initially compact and has a homogeneous density (Fig. \ref{fig:pedgroup}a). As soon as people start to interact, the model predicts an expansion of the crowd in consequence of the repulsion. At the same time, the density decreases and becomes inhomogeneous due to the anisotropy of the visual field (Fig. \ref{fig:pedgroup}b). In particular, given the orientation of the external velocity, top-bottom symmetry is preserved, because $B_r$ is symmetric with respect to $w$. However, front-rear symmetry is lost, and most people remain initially concentrated in the rear part of the group, where the influence of the mass ahead is stronger. By consequence, in this zone the velocity is lower, hence at successive times the group elongates in the horizontal direction until the distribution of people becomes again substantially homogeneous (Fig. \ref{fig:pedgroup}c). Only the motion of the leaders seems to be basically unperturbed (Figs. \ref{fig:pedgroup}b, \ref{fig:pedgroup}c), coherently with the fact that they simply follow the external velocity because nobody is in front of them.

\begin{figure}[t]
\begin{center}
\begin{minipage}[c]{0.24\textwidth}
\centering
\includegraphics[width=\textwidth,clip]{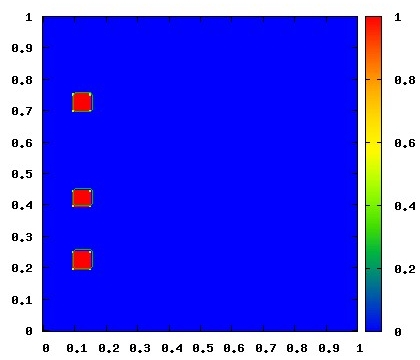} \\
(a)
\end{minipage}
\begin{minipage}[c]{0.24\textwidth}
\centering
\includegraphics[width=\textwidth,clip]{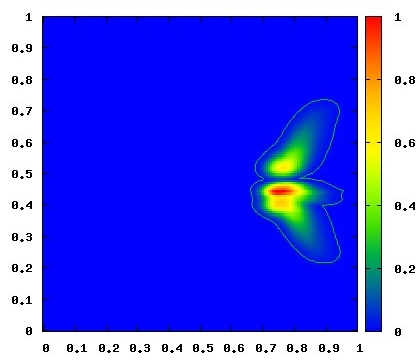} \\
(b)
\end{minipage}
\begin{minipage}[c]{0.24\textwidth}
\centering
\includegraphics[width=\textwidth,clip]{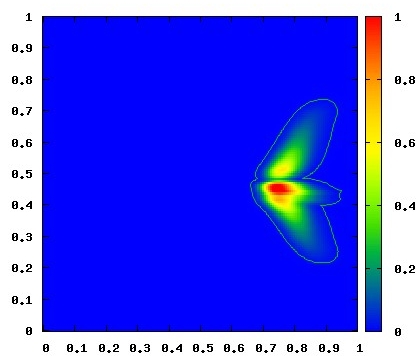} \\
(c)
\end{minipage}
\begin{minipage}[c]{0.24\textwidth}
\centering
\includegraphics[width=\textwidth,clip]{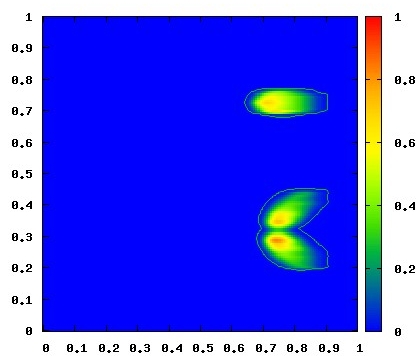} \\
(d)
\end{minipage}
\caption{Effect of cohesion in walking pedestrians. (a) Initial condition. (b) Topological cohesion makes the three clusters merge. (c) Metric cohesion with large radius $R_c^\text{max}$ gives a result qualitatively similar to the topological one. (d) Metric cohesion with small radius $R_c^\text{max}$ produces a merging only of sufficiently close clusters.}
\label{fig:cohesmacro}
\end{center}
\end{figure}

\subsubsection{Effect of cohesion}	\label{ssubsect:cohesmacro}
In this test we investigate the effect of cohesion in the macroscopic framework of pedestrians. The intelligent velocity $\nu_n$ is now as in Eq. \eqref{eq:nun-continuum}, with $F_c$ significantly greater than $F_r$ ($\vert F_c/F_r\vert =O(10^2)$) and a sensing domain for cohesion spanning the whole space around the agents ($\alpha_c=2\pi$).

The test starts with three clusters of pedestrians at the same homogeneous density, located a certain distance away from one another along the left side of the domain (Fig. \ref{fig:cohesmacro}a), which walk rightward ($w=(1,\,0)$). If $R_c^{\text{max}}$ is large enough, pedestrians are allowed to adjust the amplitude of their zone of cohesion $B_c$ so as to interact with a predefined amount of people, which in this simulation is set at $\frac{2}{3}$ of the total mass initially present in the domain (\ie, $p=\frac{2}{3}\mu_0(\Omega)$ in Eq. \eqref{eq:constraints-Bc}). Then the three clusters tend to merge in a unique group, as shown in Fig. \ref{fig:cohesmacro}b (this result can be compared with Fig. 4c in \cite{ballerini2008ira}, which shows the outcome of a similar experiment performed by a microscopic model). If instead $p=+\infty$, the zone of cohesion is fixed to its maximum size determined by $R_c^\text{max}$ (metric cohesion). In particular, for a large radius $R_c^\text{max}$ (Fig. \ref{fig:cohesmacro}c) the result is qualitatively similar to that obtained with topological cohesion, while for a small radius $R_c^\text{max}$ (Fig. \ref{fig:cohesmacro}d) only the two clusters initially sufficiently close merge, the third one being instead unaffected by the presence of other agents in the domain (see Fig. 4b in \cite{ballerini2008ira}). We refer the reader to the next Sect. \ref{ssect:commtopcorr} for comments on the importance of topological cohesion in spite of some qualitatively similar metric outcomes.

\begin{figure}[t]
\begin{center}
\begin{minipage}[c]{0.24\textwidth}
\centering
\includegraphics[width=\textwidth,clip]{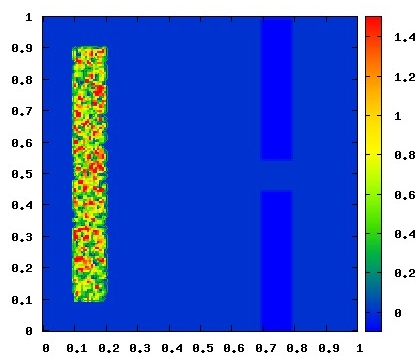} \\
(a)
\end{minipage}
\begin{minipage}[c]{0.24\textwidth}
\centering
\includegraphics[width=\textwidth,clip]{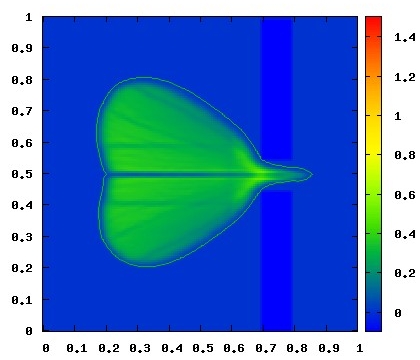} \\
(b)
\end{minipage}
\begin{minipage}[c]{0.24\textwidth}
\centering
\includegraphics[width=\textwidth,clip]{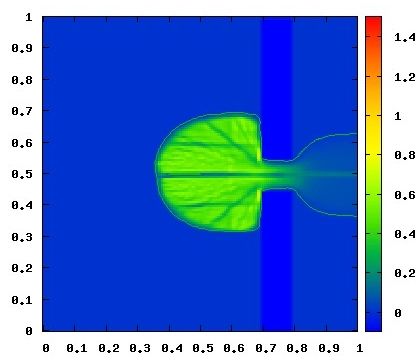} \\
(c)
\end{minipage}
\begin{minipage}[c]{0.24\textwidth}
\centering
\includegraphics[width=\textwidth,clip]{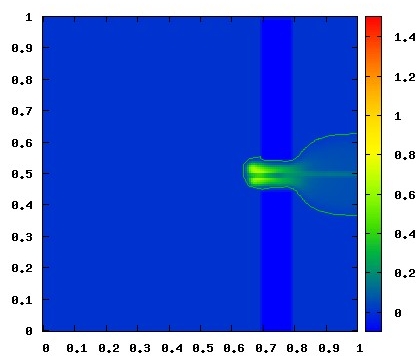} \\
(d)
\end{minipage} \\
\begin{minipage}[c]{0.24\textwidth}
\centering
\includegraphics[width=\textwidth,clip]{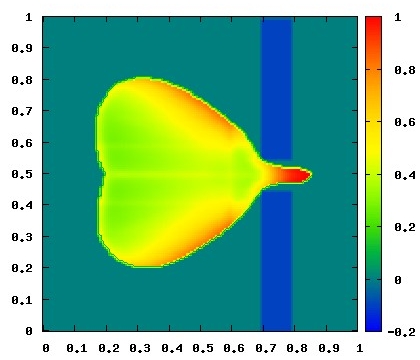} \\
(e)
\end{minipage}
\begin{minipage}[c]{0.24\textwidth}
\centering
\includegraphics[width=\textwidth,clip]{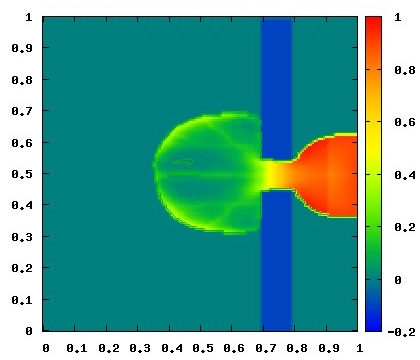} \\
(f)
\end{minipage}
\begin{minipage}[c]{0.24\textwidth}
\centering
\includegraphics[width=\textwidth,clip]{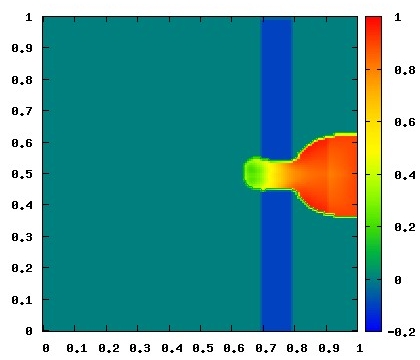} \\
(g)
\end{minipage}
\caption{Pedestrian dynamics in presence of obstacles. Upper row: a crowd wants to reach the right edge of the domain, going through a botteneck. An obstruction forms as pedestrians try to access the passage, until all people flow to the opposite side. Lower row: speed map. The speed of the crowd is inhomogeneous, with a sharp transition from low to high values across the bottleneck in correspondence of the opposite transition in the values of the density.}
\label{fig:pseudoHelb}
\end{center}
\end{figure}

\subsubsection{Dynamics in presence of obstacles}
Finally we study the motion of a crowd in a structured environment, in which some obstacles give rise to bottlenecks and direct pedestrians along preferential paths (the external velocity field is no longer homogeneous in space). In particular, we consider the case of a group of pedestrians wanting to go through a narrow passage, obtained by placing two obstacles in front of each other as in Fig. \ref{fig:pseudoHelb}a. The external velocity $w$ is obtained by solving Laplace's equation for the potential $u$, along with sliding boundary conditions at the obstacle edges (Neumann conditions). A Dirichlet boundary condition is instead imposed on the right edge of the domain, in order to set the potential at its maximum and to identify pedestrians' target. In this simulation cohesion is not active ($F_c=0$), while repulsion is felt more strongly than in case of motion in clear environments ($F_r=-10$).

Starting from an inhomogeneous density of people, confined in the left area of the domain $\Omega$ (Fig. \ref{fig:pseudoHelb}a), the self-organization of the crowd predicted by the model is as follows. When approaching the bottleneck (Fig. \ref{fig:pseudoHelb}b), people initially pass through at the maximum speed (Fig. \ref{fig:pseudoHelb}e), however not all pedestrians can access the bottleneck at the same time and an obstruction forms (Fig. \ref{fig:pseudoHelb}c). Speed before the bottleneck is low, some individuals in the middle of the group are even forced to stop, whereas behind the bottleneck it attains again its maximum (Fig. \ref{fig:pseudoHelb}f). After a certain time, the whole group flows through the bottleneck and the obstruction is depleted (Figs. \ref{fig:pseudoHelb}d, \ref{fig:pseudoHelb}g). This simulation compares qualitatively well with that proposed in \cite{HeFaMoVi} by means of a microscopic model.

\begin{figure}[t]
\begin{center}
\includegraphics[width=0.4\textwidth]{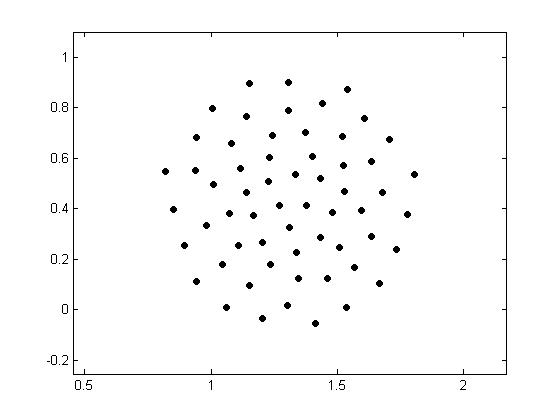}
\caption{Two-dimensional globular clusters, metric interaction.}
\label{fig:nocrystal}
\end{center}
\end{figure}

\begin{figure}[t]
\begin{center}
\begin{minipage}[c]{0.32\textwidth}
\centering
\includegraphics[width=\textwidth,clip]{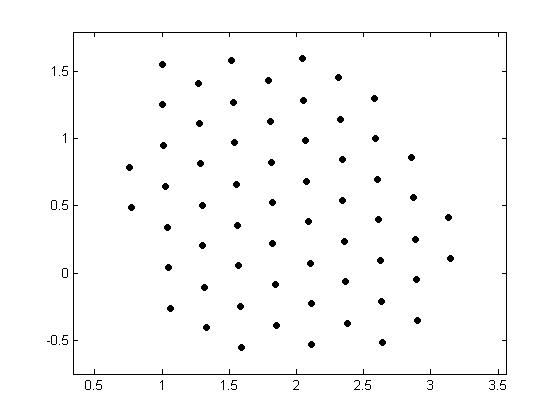}\\
(a)
\end{minipage}
\begin{minipage}[c]{0.32\textwidth}
\centering
\includegraphics[width=\textwidth,clip]{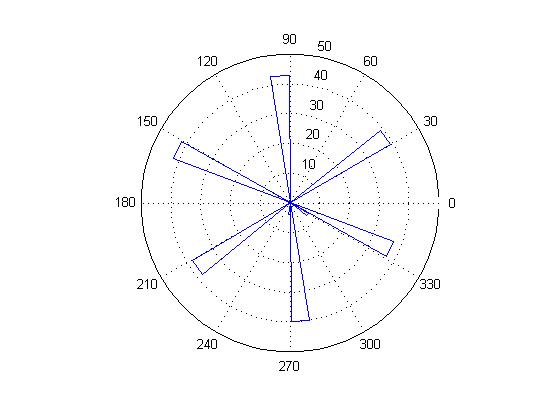}\\
(b)
\end{minipage}
\begin{minipage}[c]{0.32\textwidth}
\centering
\includegraphics[width=\textwidth,clip]{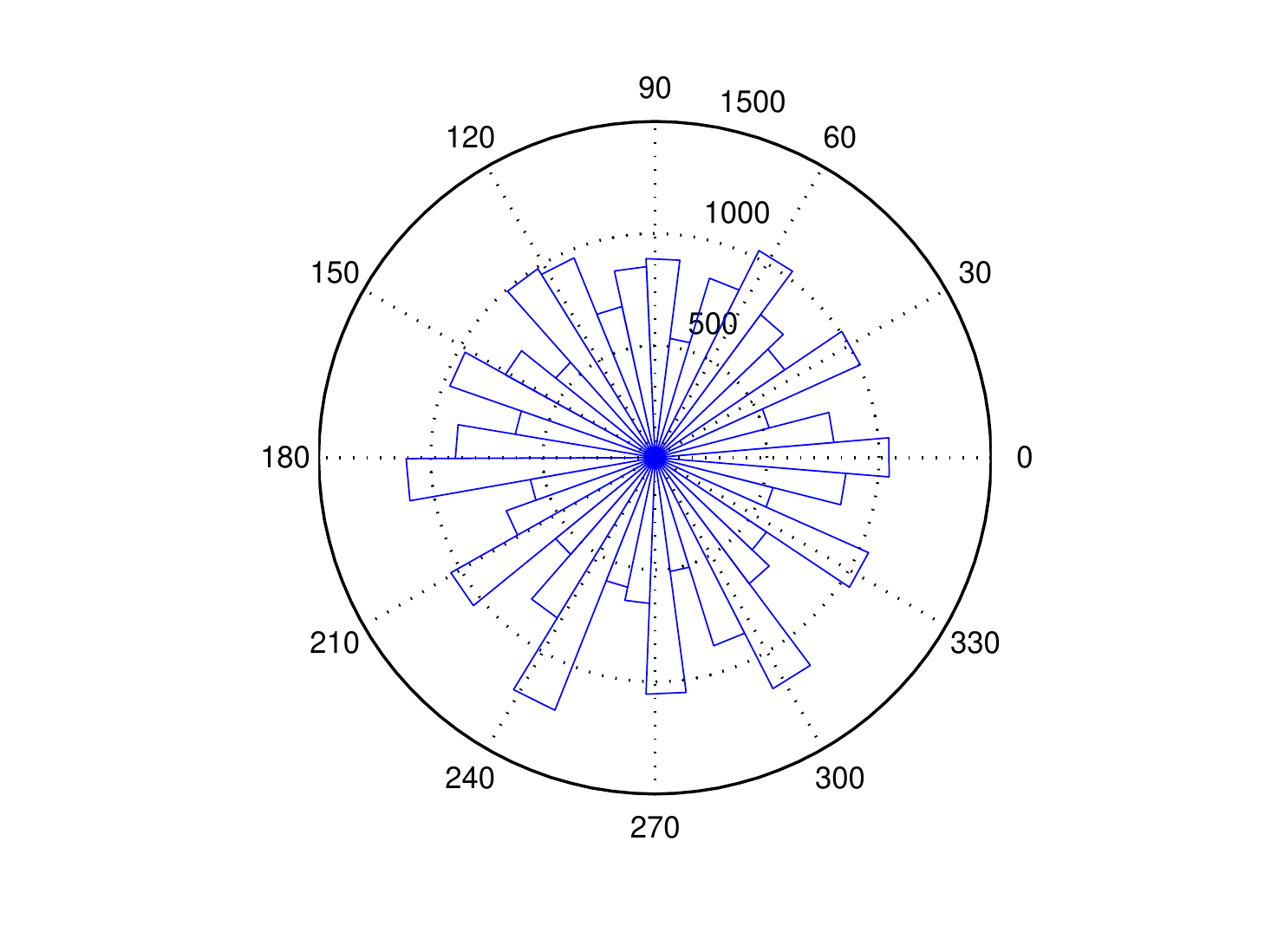}\\
(c)
\end{minipage}
\caption{Two-dimensional globular clusters, topological interaction. (a) Typical outcome with crystal structure, (b) angle distribution among the agents for 1 run, and (c) angle distribution among the agents for 100 runs.}
\label{fig:crystal}
\end{center}
\end{figure}

\subsection{Microscopic self-organization in animals}	\label{sect:num-micro}
For animal groups, we consider the complete structure of the intelligent velocity, in which both cohesion and repulsion are active. We investigate the importance of the topological correction and the effects of the anisotropic interaction, varying $p$ and the angles $\alpha_r$ and $\alpha_c$. We choose $R_r$ as few times the body size $\ell$ of the agents, while we allow for a very large maximum radius $R_c^{\text{max}}$ for cohesion.

\subsubsection{Two-dimensional globular cluster}
For the first test we set $\alpha_r=\alpha_c=2\pi$ and $p=N$. The choice $p=N$, together with a large $R_c^{\text{max}}$, implies that cohesion is basically metric and all-to-all. The parameters $F_c$ and $F_r$ are of the same order of magnitude. The system reaches a stable equilibrium in few iterations, forming a ball-like group with an irregular internal structure. In Fig. \ref{fig:nocrystal} we show the typical outcome. If we introduce the topological correction, choosing $p=7$, the system reaches again a stable equilibrium in few iterations, but this time a group forms, in which all IPs are at the same distance from each other. Every internal IP is surrounded by six group mates forming an hexagon, in a crystal-like structure (Fig. \ref{fig:crystal}a) (cf. also \cite{gregoire2003mst,li2008mms}).

By computing the distribution of the angles between each IP and its neighbors, we investigate the orientation of the hexagons. In Figs. \ref{fig:crystal}b, \ref{fig:crystal}c we show the angle distribution for 1 run and 100 runs, whence we can deduce that the orientation of the hexagons is mainly random. 

\begin{figure}[t]
\begin{center}
\begin{minipage}[c]{0.45\textwidth}
\centering
\includegraphics[width=\textwidth]{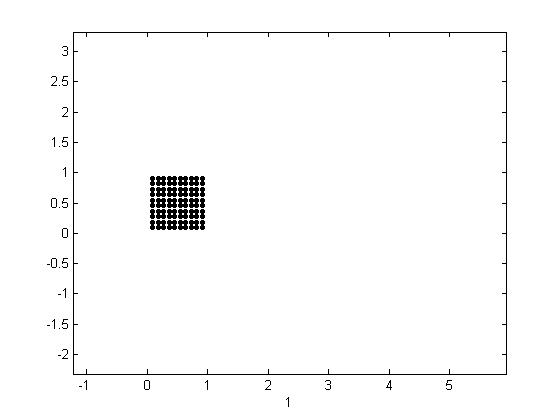}\\
(a)
\end{minipage}
\begin{minipage}[c]{0.45\textwidth}
\centering
\includegraphics[width=\textwidth]{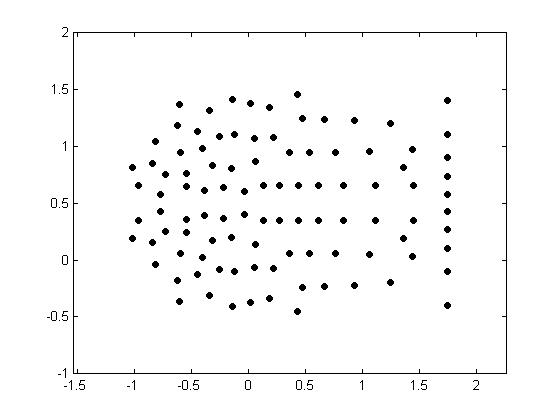}\\
(b)
\end{minipage}
\caption{Macroscopic and microscopic models can reproduce the same patterns, confirming that they come from the same modeling framework.}
\label{fig:esempiocomune1}
\end{center}
\end{figure}

By switching cohesion off ($F_c=0$) and setting up a mild repulsion among the agents ($F_r=-0.05$) with a frontal visual field only ($\alpha_r=\pi$), we can also mimic with the microscopic model the spontaneous arrangement of a group of IPs described in Sect. \ref{ssubsect:spontarrang} by the macroscopic model. Starting from a regular square configuration (Fig. \ref{fig:esempiocomune1}a), the IPs move rightward and interact only with the IPs in front of them. In Fig. \ref{fig:esempiocomune1}b we show the final stable configuration, directly comparable with that in Fig. \ref{fig:pedgroup}c.

\begin{figure}[t]
\begin{center}
\begin{minipage}[c]{0.32\textwidth}
\centering
\includegraphics[width=\textwidth]{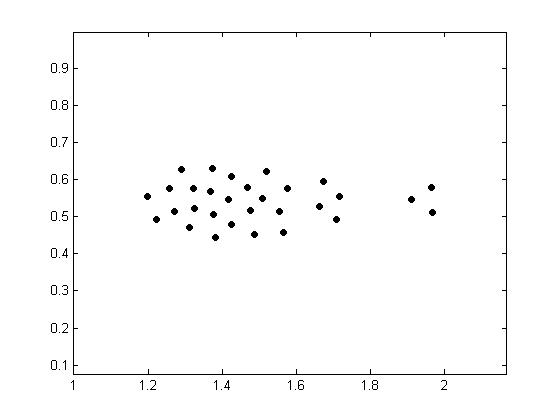}\\
(a)
\end{minipage}
\begin{minipage}[c]{0.32\textwidth}
\centering
\includegraphics[width=\textwidth]{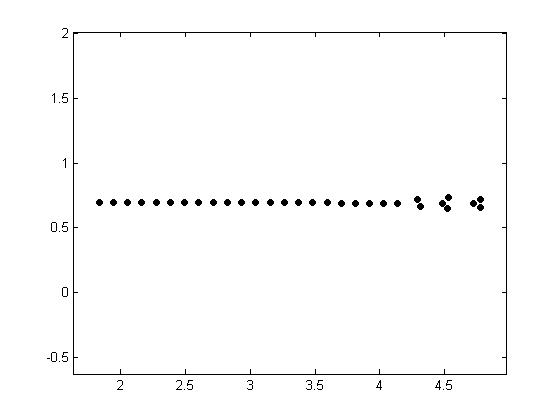}\\
(b)
\end{minipage}
\begin{minipage}[c]{0.32\textwidth}
\centering
\includegraphics[width=\textwidth]{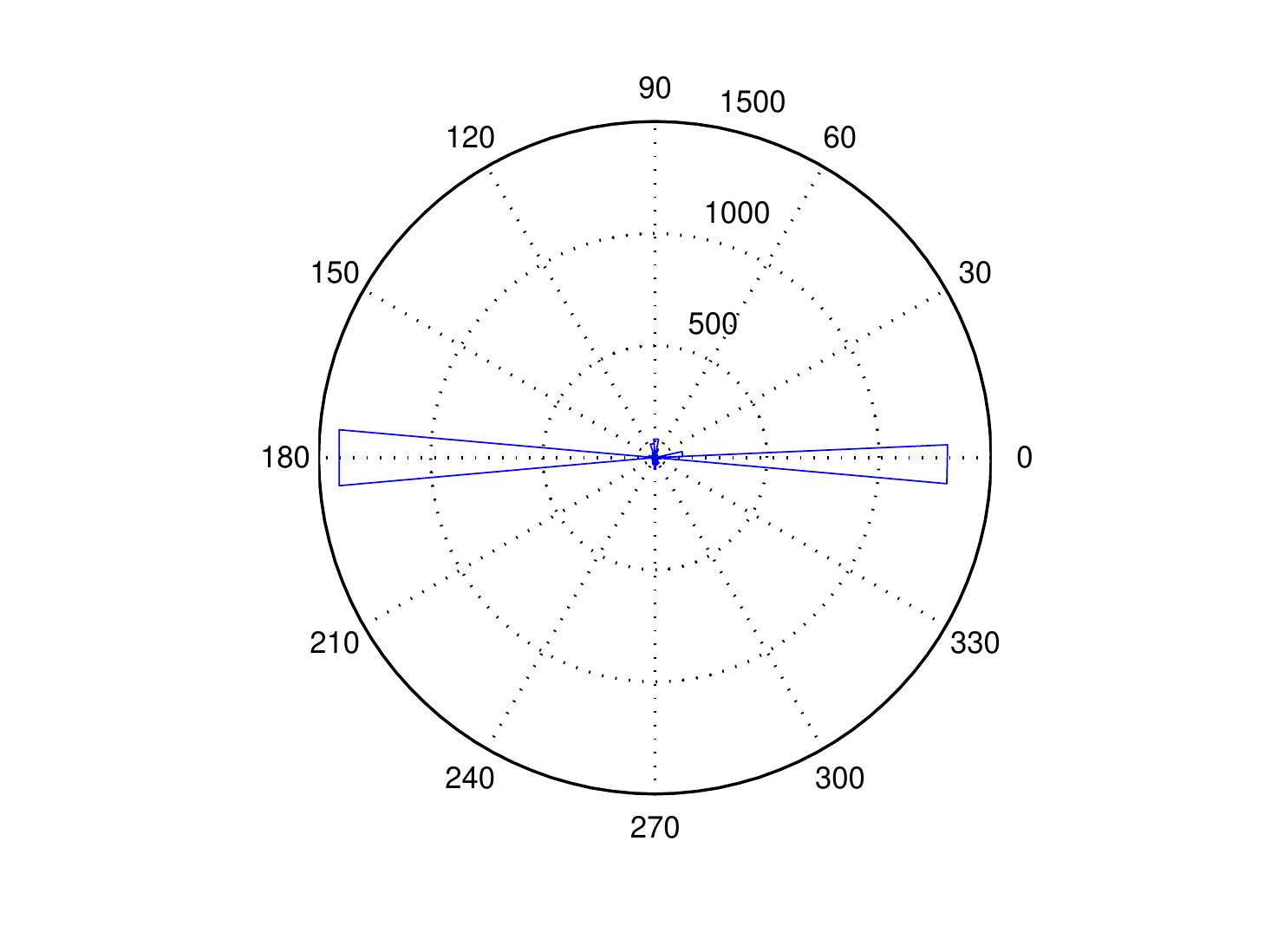}\\
(c)
\end{minipage}
\caption{Line formations: (a) purely metric interaction, (b) topological interaction, and (c) angle distribution among the agents for 100 runs.}
\label{fig:line}
\end{center}
\end{figure}

\subsubsection{Line formation}
Here we set $\alpha_r=\frac{\pi}{4}$, $\alpha_c=\pi$, and $p=N$. Cohesion is now greater than repulsion ($F_c>\vert F_r\vert$). The system does not reach an equilibrium (see Fig. \ref{fig:line}a), nevertheless the resulting pattern is strongly different from that of the previous test. The new outcome is mainly due to the modification in the in the relative magnitude of $F_c$ and $F_r$ and to different angles $\alpha_c$, $\alpha_r$, which imply a different anisotropy in the sensing zones.

When introducing the topological correction with $p=7$, the system reaches an equilibrium after few hundreds of iterations, forming a line oriented in the direction of the motion. Neglecting small contributions of the repulsion component we obtain a rather stable line (but for some little border effect in the head due to that group leaders cannot interact with $p$ group mates ahead), see Fig. \ref{fig:line}b. Lines are an example of pattern produced by self-organization of terrestrial animals like migrating penguins or elephants.

A statistic on the distribution of the angles among the agents for $100$ runs (Fig. \ref{fig:line}c) shows that a line configuration is always reached.

In \cite{cristiani2009eac-preprint} the authors show the effect of a continuous variation of the pair $(\alpha_r,\,\alpha_c)$ from $(2\pi,\,2\pi)$ to $(\frac{\pi}{4},\,\pi)$. As we have seen in the previous test, the first choice corresponds to a two-dimensional globular cluster. They find that wide angles induce a stretching of the group along the vertical direction, although in most runs the system does not reach an equilibrium. Conversely, small angles lead to a strong elongation of the group in the horizontal direction, and, in the limit case, to the formation of a line.

\subsection{Comments on the effect of topological correction}	\label{ssect:commtopcorr}
The previous tests clearly show that topological cohesion between IPs greatly changes the resulting pattern. This does not mean that it is impossible to obtain similar structures with a purely metric cohesion, by duly tuning the radius of cohesion (see e.g., the macroscopic test in Sect. \ref{ssubsect:cohesmacro}). However, the topological correction is essential in order to deal with a large value of the maximum radius allowed. Indeed, as we have recalled in Sects. \ref{sect:class-vs-int}, \ref{sect:model}, a metric upper bound $R_c^\text{max}$ to the radius of cohesion $R_c$ exists, which translates the fact that IPs are in no case concerned with very far mates, and which should necessarily coincide with the fixed radius of cohesion in a purely metric approach. Now, $R_c^\text{max}$ is in general rather large, because IPs are able to see quite far, and can be attracted even by far fellows if necessary. By consequence, once the group is formed, a purely metric cohesion with a large $R_c^\text{max}$ would imply attraction with an unreasonable number of other IPs, instead of feeling comfortable with the proper amount of IPs in the surroundings. Thus the topological correction is the only way to stay cohesive with a reasonable number of group mates while keeping a large maximum radius $R_c^\text{max}$.

As a further confirmation of this, the test on the effect of cohesion in the macroscopic model (cf. Sect. \ref{ssubsect:cohesmacro}) shows that a small value of $R_c^\text{max}$ in the purely metric approach distorts the cohesion itself: aggregation is only partial, since little far group mates may not be seen.

\section{Conclusions and research perspectives}	\label{sect:conclusions}
In this paper we have introduced a modeling framework for self-organizing intelligent particles, which takes into account both macroscopic and microscopic points of view, and is suitable to include topological and anisotropic interactions in an easy way and in any dimension.

The results we have obtained from our numerical simulations suggest that these two features alone let self-organization emerge spontaneously, without forcing pattern formation via \emph{ad hoc} agent-specific behavioral rules. In this respect, the most important parameters of the model are the ratio $\vert F_c/F_r\vert$, \ie, the relative strength of cohesive vs. repulsive terms, and the angles $\alpha_c,\,\alpha_r$, in other words the span of the sensing zones for cohesion and repulsion, respectively.

The modeling technique by time-evolving measures that we have introduced is promising and deserves further investigation, because it enables one to address macroscopic and microscopic modeling by common mathematical structures and tools. Specifically, we have used the macroscopic scale to study self-organization in pedestrian flows, as in this case clearly distinguishable patterns emerge only when the density of people is sufficiently large. In addition, the macroscopic approach may be profitably used in control and optimization problems connected to the improvement of the flow and the safety of crowds. For instance, as recalled in Sect. \ref{sect:self-org} about Braess' paradox, this might imply optimization of the locations of some obstacles in crowded environments, like train stations or shopping malls. Conversely, we have studied self-organization in animals at the microscopic scale in order to catch the fine internal structure of the group and to highlight the appearance of regular structures (e.g., crystals) formed by few agents. The microscopic approach may be used to address problems in which the granularity plays an essential role. Furthermore, it may allow to introduce in the behavior of the agents stochastic effects which are not suited to an averaged macroscopic framework.

\section*{Acknowledgements}
Credits for pictures of Fig. \ref{fig:photos}:
(a), (b) Copyright Dirk Helbing. (c) Copyright Bjarne Winkler\footnote{\texttt{http://epod.typepad.com/blog/2006/06/black-sun-in-denmark.html}}.
(d) Copyright Ryan Lukeman\footnote{\texttt{http://www.iam.ubc.ca/\~{}lukeman.html}}.
(e) Screenshot from the video ``African elephants walks the savanna grassland'' by Joseph Kimojino\footnote{\texttt{http://vimeo.com}}.

This research was partially supported by the Network of Excellence project HYCON (2004-2009) and by the FIRB 2005 research project CASHMA.

A. Tosin further acknowledges the support of a fellowship by the National Institute for Advanced Mathematics (INdAM) and the ``Compagnia di San Paolo'' foundation.

\end{document}